\documentclass[10pt,a4paper,onecolumn]{article}
\usepackage{marginnote}
\usepackage{graphicx}
\usepackage{xcolor}
\usepackage{authblk,etoolbox}
\usepackage{titlesec}
\usepackage{calc}
\usepackage{tikz}
\usepackage{hyperref}
\hypersetup{colorlinks,breaklinks,
            urlcolor=[rgb]{0.0, 0.5, 1.0},
            linkcolor=[rgb]{0.0, 0.5, 1.0}}
\usepackage{caption}
\usepackage{tcolorbox}
\usepackage{amssymb,amsmath}
\usepackage{ifxetex,ifluatex}
\usepackage{seqsplit}
\usepackage{xstring}

\usepackage{float}
\let\origfigure\figure
\let\endorigfigure\endfigure
\renewenvironment{figure}[1][2] {
    \expandafter\origfigure\expandafter[H]
} {
    \endorigfigure
}

\usepackage{fixltx2e} % provides \textsubscript
\usepackage[
  backend=biber,
]{biblatex}
\bibliography{references.bib}

\let\textttOrig=\texttt
\def\texttt#1{\expandafter\textttOrig{\seqsplit{#1}}}
\renewcommand{\seqinsert}{\ifmmode
  \allowbreak
  \else\penalty6000\hspace{0pt plus 0.02em}\fi}

\makeatletter
\let\href@Orig=\href
\def\href@Urllike#1#2{\href@Orig{#1}{\begingroup
    \def\Url@String{#2}\Url@FormatString
    \endgroup}}
\def\href@Notdoi#1#2{\def\tempa{#1}\def\tempb{#2}%
  \ifx\tempa\tempb\relax\href@Urllike{#1}{#2}\else
  \href@Orig{#1}{#2}\fi}
\def\href#1#2{%
  \IfBeginWith{#1}{https://doi.org}%
  {\href@Urllike{#1}{#2}}{\href@Notdoi{#1}{#2}}}
\makeatother

\newlength{\cslhangindent}
\newlength{\csllabelwidth}
\newenvironment{CSLReferences}[3] % #1 hanging-ident, #2 entry spacing
 {% don't indent paragraphs
  \setlength{\parindent}{0pt}
  \ifodd #1 \everypar{\setlength{\hangindent}{\cslhangindent}}\ignorespaces\fi
  \ifnum #2 > 0
  \setlength{\parskip}{#2\baselineskip}
  \fi
 }%
 {}
\usepackage{calc}

\usepackage[top=3.5cm, bottom=3cm, right=1.5cm, left=1.0cm,
            headheight=2.2cm, reversemp, includemp, marginparwidth=4.5cm]{geometry}

\titleformat{\section}
  {\normalfont\sffamily\Large\bfseries}
  {}{0pt}{}
\titleformat{\subsection}
  {\normalfont\sffamily\large\bfseries}
  {}{0pt}{}
\titleformat{\subsubsection}
  {\normalfont\sffamily\bfseries}
  {}{0pt}{}
\titleformat*{\paragraph}
  {\sffamily\normalsize}

\usepackage{fancyhdr}
\makeatletter
\let\ps@plain\ps@fancy
\fancyheadoffset[L]{4.5cm}
\fancyfootoffset[L]{4.5cm}

\definecolor{linky}{rgb}{0.0, 0.5, 1.0}

\newtcolorbox{repobox}
   {colback=red, colframe=red!75!black,
     boxrule=0.5pt, arc=2pt, left=6pt, right=6pt, top=3pt, bottom=3pt}

\newcommand{\ExternalLink}{%
   \tikz[x=1.2ex, y=1.2ex, baseline=-0.05ex]{%
       \begin{scope}[x=1ex, y=1ex]
           \clip (-0.1,-0.1)
               --++ (-0, 1.2)
               --++ (0.6, 0)
               --++ (0, -0.6)
               --++ (0.6, 0)
               --++ (0, -1);
           \path[draw,
               line width = 0.5,
               rounded corners=0.5]
               (0,0) rectangle (1,1);
       \end{scope}
       \path[draw, line width = 0.5] (0.5, 0.5)
           -- (1, 1);
       \path[draw, line width = 0.5] (0.6, 1)
           -- (1, 1) -- (1, 0.6);
       }
   }

\patchcmd{\@maketitle}{center}{flushleft}{}{}
\patchcmd{\@maketitle}{center}{flushleft}{}{}
\patchcmd{\@maketitle}{\LARGE}{\LARGE\sffamily}{}{}
\def\maketitle{{%
  
  \AB@maketitle}}
\makeatletter
\renewcommand\AB@affilsepx{ \protect\Affilfont}
\renewcommand\AB@affilnote[1]{{\bfseries #1}\hspace{3pt}}
\renewcommand{\affil}[2][]%
   {\newaffiltrue\let\AB@blk@and\AB@pand
      \if\relax#1\relax\def\AB@note{\AB@thenote}\else\def\AB@note{#1}%
        \setcounter{Maxaffil}{0}\fi
        \begingroup
        \let\href=\href@Orig
        \let\texttt=\textttOrig
        \let\protect\@unexpandable@protect
        \def\thanks{\protect\thanks}\def\footnote{\protect\footnote}%
        \@temptokena=\expandafter{\AB@authors}%
        {\def\\{\protect\\\protect\Affilfont}\xdef\AB@temp{#2}}%
         \xdef\AB@authors{\the\@temptokena\AB@las\AB@au@str
         \protect\\[\affilsep]\protect\Affilfont\AB@temp}%
         \gdef\AB@las{}\gdef\AB@au@str{}%
        {\def\\{, \ignorespaces}\xdef\AB@temp{#2}}%
        \@temptokena=\expandafter{\AB@affillist}%
        \xdef\AB@affillist{\the\@temptokena \AB@affilsep
          \AB@affilnote{\AB@note}\protect\Affilfont\AB@temp}%
      \endgroup
       \let\AB@affilsep\AB@affilsepx
}
\makeatother

\renewcommand\Affilfont{\sffamily\small\mdseries}
\ifnum 0\ifxetex 1\fi\ifluatex 1\fi=0 % if pdftex
  \usepackage[T1]{fontenc}
  \usepackage[utf8]{inputenc}

\else % if luatex or xelatex
  \ifxetex
    \usepackage{mathspec}
  \else
    \usepackage{fontspec}
  \fi
  \defaultfontfeatures{Ligatures=TeX,Scale=MatchLowercase}

\fi
\IfFileExists{upquote.sty}{\usepackage{upquote}}{}
\IfFileExists{microtype.sty}{%
\usepackage{microtype}
\UseMicrotypeSet[protrusion]{basicmath} % disable protrusion for tt fonts
}{}

\usepackage{hyperref}
\hypersetup{unicode=true,
            pdftitle={The Data Behind Dark Matter: Exploring Galactic Rotation},
            pdfborder={0 0 0},
            breaklinks=true}
\usepackage{longtable,booktabs}

\let\addcontentslineOrig=\addcontentsline
\def\addcontentsline#1#2#3{\bgroup
  \let\texttt=\textttOrig\addcontentslineOrig{#1}{#2}{#3}\egroup}
\let\markbothOrig\markboth
\def\markboth#1#2{\bgroup
  \let\texttt=\textttOrig\markbothOrig{#1}{#2}\egroup}
\let\markrightOrig\markright
\def\markright#1{\bgroup
  \let\texttt=\textttOrig\markrightOrig{#1}\egroup}

\IfFileExists{parskip.sty}{%
\usepackage{parskip}
}{% else
\setlength{\parindent}{0pt}
\setlength{\parskip}{6pt plus 2pt minus 1pt}
}
\providecommand{\tightlist}{%
  \setlength{\itemsep}{0pt}\setlength{\parskip}{0pt}}
\ifx\paragraph\undefined\else
\let\oldparagraph\paragraph
\renewcommand{\paragraph}[1]{\oldparagraph{#1}\mbox{}}
\fi
\ifx\subparagraph\undefined\else
\let\oldsubparagraph\subparagraph
\renewcommand{\subparagraph}[1]{\oldsubparagraph{#1}\mbox{}}
\fi

\title{The Data Behind Dark Matter: Exploring Galactic Rotation}

        \author[1]{A.N. Villano}
          \author[2]{Kitty C. Harris}
          \author[3]{Judit Bergfalk}
          \author[1]{Raphael Hatami}
          \author[4]{Francis Vititoe}
          \author[3]{Julia Johnston}
    
      \affil[1]{Department of Physics, University of Colorado Denver,
Denver CO 80217, USA}
      \affil[2]{Integrated Sciences, University of Colorado Denver,
Denver CO 80217, USA}
      \affil[3]{Astrophysical \& Planetary Sciences, University of
Colorado Boulder, Boulder, CO 80309, USA}
      \affil[4]{Department of Physics, University of Colorado Boulder,
Boulder, CO 80309, USA}
  \date{\vspace{-5ex}}

\begin{document}
\maketitle

\marginpar{
  %\hrule
  \sffamily\small

  {\bfseries DOI:} \href{https://doi.org/DOI unavailable}{\color{linky}{DOI unavailable}}

  \vspace{2mm}

  {\bfseries Software}
  \begin{itemize}
    \setlength\itemsep{0em}
    \item \href{N/A}{\color{linky}{Review}} \ExternalLink
    \item \href{NO_REPOSITORY}{\color{linky}{Repository}} \ExternalLink
    \item \href{DOI unavailable}{\color{linky}{Archive}} \ExternalLink
  \end{itemize}

  \vspace{2mm}

  {\bfseries Submitted:} N/A\\
  {\bfseries Published:} N/A

  \vspace{2mm}
  {\bfseries License}\\
  Authors of papers retain copyright and release the work under a Creative Commons Attribution 4.0 International License (\href{http://creativecommons.org/licenses/by/4.0/}{\color{linky}{CC BY 4.0}}).
}

\hypertarget{summary}{%
\section{Summary}\label{summary}}

By analyzing the rotational velocities of bodies in galaxies, physicists
and astronomers have found that there seems to be something missing in
our understanding of these galaxies. One theory is that there is some
invisible matter present that does not interact with light --- that is,
these galaxies contain dark matter (Rubin et al., 1978).

Participants in this workshop have the opportunity to explore dark
matter through scientific literature-based (Fraternali, F. et al., 2011;
Jimenez et al., 2003; Karukes, E. V. et al., 2015; Naray et al., 2008;
Richards et al., 2015) galactic rotation curves both by using
interactive programs and by editing Python code. This will give
participants an understanding of how physicists arrived at the idea of
dark matter, showing them the difference between curve fits with and
without dark matter components. Understanding dark matter's
epistemological origins will help participants formulate their own
opinions on the dark matter debate.

\hypertarget{materials}{%
\subsection{Materials}\label{materials}}

This project consists of several modules in the form of Jupyter
notebooks (Kluyver et al., 2016):

\begin{longtable}[]{@{}ll@{}}
\toprule
\begin{minipage}[b]{(\columnwidth - 1\tabcolsep) * \real{0.68}}\raggedright
\textbf{File Name}\strut
\end{minipage} &
\begin{minipage}[b]{(\columnwidth - 1\tabcolsep) * \real{0.32}}\raggedright
Description\strut
\end{minipage}\tabularnewline
\midrule
\endhead
\begin{minipage}[t]{(\columnwidth - 1\tabcolsep) * \real{0.68}}\raggedright
\textbf{01\_DM\_Rotation\_Curve\_Intro.ipynb}\strut
\end{minipage} &
\begin{minipage}[t]{(\columnwidth - 1\tabcolsep) * \real{0.32}}\raggedright
Animations and rotation curve plots demonstrating three types of
rotational motion.\strut
\end{minipage}\tabularnewline
\begin{minipage}[t]{(\columnwidth - 1\tabcolsep) * \real{0.68}}\raggedright
\strut
\end{minipage} &
\begin{minipage}[t]{(\columnwidth - 1\tabcolsep) * \real{0.32}}\raggedright
\strut
\end{minipage}\tabularnewline
\begin{minipage}[t]{(\columnwidth - 1\tabcolsep) * \real{0.68}}\raggedright
\textbf{02\_Widget\_NGC5533\_DMonly.ipynb}\strut
\end{minipage} &
\begin{minipage}[t]{(\columnwidth - 1\tabcolsep) * \real{0.32}}\raggedright
Interactive widget to introduce dark matter.\strut
\end{minipage}\tabularnewline
\begin{minipage}[t]{(\columnwidth - 1\tabcolsep) * \real{0.68}}\raggedright
\strut
\end{minipage} &
\begin{minipage}[t]{(\columnwidth - 1\tabcolsep) * \real{0.32}}\raggedright
\strut
\end{minipage}\tabularnewline
\begin{minipage}[t]{(\columnwidth - 1\tabcolsep) * \real{0.68}}\raggedright
\textbf{03\_Measured\_Data\_Plotting.ipynb}\strut
\end{minipage} &
\begin{minipage}[t]{(\columnwidth - 1\tabcolsep) * \real{0.32}}\raggedright
Rotation curve plotting of measured velocities to visualize star and gas
motions in a galaxy.\strut
\end{minipage}\tabularnewline
\begin{minipage}[t]{(\columnwidth - 1\tabcolsep) * \real{0.68}}\raggedright
\strut
\end{minipage} &
\begin{minipage}[t]{(\columnwidth - 1\tabcolsep) * \real{0.32}}\raggedright
\strut
\end{minipage}\tabularnewline
\begin{minipage}[t]{(\columnwidth - 1\tabcolsep) * \real{0.68}}\raggedright
\textbf{04\_Plotting\_Rotation\_Curves.ipynb}\strut
\end{minipage} &
\begin{minipage}[t]{(\columnwidth - 1\tabcolsep) * \real{0.32}}\raggedright
Plotting the rotation curves of galaxy components.\strut
\end{minipage}\tabularnewline
\begin{minipage}[t]{(\columnwidth - 1\tabcolsep) * \real{0.68}}\raggedright
\strut
\end{minipage} &
\begin{minipage}[t]{(\columnwidth - 1\tabcolsep) * \real{0.32}}\raggedright
\strut
\end{minipage}\tabularnewline
\begin{minipage}[t]{(\columnwidth - 1\tabcolsep) * \real{0.68}}\raggedright
\textbf{05\_Widget\_NGC5533\_All\_Components.ipynb}\strut
\end{minipage} &
\begin{minipage}[t]{(\columnwidth - 1\tabcolsep) * \real{0.32}}\raggedright
Interactive widget to visualize the components of the galaxy NGC
5533.\strut
\end{minipage}\tabularnewline
\begin{minipage}[t]{(\columnwidth - 1\tabcolsep) * \real{0.68}}\raggedright
\strut
\end{minipage} &
\begin{minipage}[t]{(\columnwidth - 1\tabcolsep) * \real{0.32}}\raggedright
\strut
\end{minipage}\tabularnewline
\begin{minipage}[t]{(\columnwidth - 1\tabcolsep) * \real{0.68}}\raggedright
\textbf{06\_Plotting\_SPARC\_Data.ipynb}\strut
\end{minipage} &
\begin{minipage}[t]{(\columnwidth - 1\tabcolsep) * \real{0.32}}\raggedright
Plotting the components of galactic rotation curves using the SPARC
database of 175 galaxies.\strut
\end{minipage}\tabularnewline
\begin{minipage}[t]{(\columnwidth - 1\tabcolsep) * \real{0.68}}\raggedright
\strut
\end{minipage} &
\begin{minipage}[t]{(\columnwidth - 1\tabcolsep) * \real{0.32}}\raggedright
\strut
\end{minipage}\tabularnewline
\begin{minipage}[t]{(\columnwidth - 1\tabcolsep) * \real{0.68}}\raggedright
\textbf{07\_Bonus\_Bulge\_Rotation\_Curve.ipynb}\strut
\end{minipage} &
\begin{minipage}[t]{(\columnwidth - 1\tabcolsep) * \real{0.32}}\raggedright
Constructing a rotation curve for the bulge component using
empirically-derived parameters.\strut
\end{minipage}\tabularnewline
\begin{minipage}[t]{(\columnwidth - 1\tabcolsep) * \real{0.68}}\raggedright
\strut
\end{minipage} &
\begin{minipage}[t]{(\columnwidth - 1\tabcolsep) * \real{0.32}}\raggedright
\strut
\end{minipage}\tabularnewline
\begin{minipage}[t]{(\columnwidth - 1\tabcolsep) * \real{0.68}}\raggedright
\textbf{08\_Interactive\_Fitting.ipynb}\strut
\end{minipage} &
\begin{minipage}[t]{(\columnwidth - 1\tabcolsep) * \real{0.32}}\raggedright
Interactive curve fitting.\strut
\end{minipage}\tabularnewline
\begin{minipage}[t]{(\columnwidth - 1\tabcolsep) * \real{0.68}}\raggedright
\strut
\end{minipage} &
\begin{minipage}[t]{(\columnwidth - 1\tabcolsep) * \real{0.32}}\raggedright
\strut
\end{minipage}\tabularnewline
\begin{minipage}[t]{(\columnwidth - 1\tabcolsep) * \real{0.68}}\raggedright
\textbf{09\_Widget\_SPARC\_Galaxies.ipynb}\strut
\end{minipage} &
\begin{minipage}[t]{(\columnwidth - 1\tabcolsep) * \real{0.32}}\raggedright
Interactive widget to visualize the components of multiple galaxies
using the SPARC database.\strut
\end{minipage}\tabularnewline
\begin{minipage}[t]{(\columnwidth - 1\tabcolsep) * \real{0.68}}\raggedright
\strut
\end{minipage} &
\begin{minipage}[t]{(\columnwidth - 1\tabcolsep) * \real{0.32}}\raggedright
\strut
\end{minipage}\tabularnewline
\begin{minipage}[t]{(\columnwidth - 1\tabcolsep) * \real{0.68}}\raggedright
\textbf{10\_Bonus\_Black\_Holes\_as\_DM.ipynb}\strut
\end{minipage} &
\begin{minipage}[t]{(\columnwidth - 1\tabcolsep) * \real{0.32}}\raggedright
Considering tiny black holes as dark matter candidates.\strut
\end{minipage}\tabularnewline
\bottomrule
\end{longtable}

\hypertarget{statement-of-need}{%
\section{Statement of Need}\label{statement-of-need}}

The primary goal of our project is to present rotation curve development
and research in a versatile and approachable format for anyone to
explore, learn from, and build upon. Rotation curves are a key empirical
artifact through which dark matter can be observed and analyzed (Rubin
et al., 1978); however, a thorough, start-to-finish description of the
rotation curve building process is typically not given in scientific
publications. Furthermore, software tools used in rotation curve
literature are generally difficult for inexperienced users; for example,
the GIPSY software package is very thorough but does not provide any
introduction as it is intended for experienced users with a firm grasp
on rotation curve components (Kapteyn Astronomical Institute, 1992).
Therefore, a rigorous yet accessible learning module is needed to
provide an entry point for any individual interested in investigating
the effect of dark matter in spiral galaxies. Our workshop is designed
to present a convenient platform for developing basic rotation curves
focused on introducing newcomers to the concepts necessary for
understanding galactic rotation. This is achieved by leading users
through hands-on computational activities, including building and
plotting their own rotation curves.

\hypertarget{learning-objectives}{%
\section{Learning Objectives}\label{learning-objectives}}

The learning objectives for these modules are:

\begin{enumerate}
\def\labelenumi{\arabic{enumi}.}
\tightlist
\item
  Provide a working space where people can connect with current
  literature and identify as scientists.
\item
  Educate curious students or other individuals on the basic concepts of
  rotation curves, as related to the current problems and mysteries
  regarding dark matter in the universe.
\item
  Provide users with accessible activities relating to the basic
  principles of rotation curve composition. This includes:

  \begin{enumerate}
  \def\labelenumii{\alph{enumii}.}
  \tightlist
  \item
    facilitating the introduction of rotation curve concepts via
    open-source code.
  \item
    interactive programs to provide users with practical and tangible
    approach of what producing rotation curves involves.
  \end{enumerate}
\item
  Understand data and models by interacting directly with equations and
  figures.
\end{enumerate}

Most of the content provided in these modules has been presented and
taught in previous workshops/research symposiums (University of Colorado
Denver: Data Science Symposium 2021 (Villano et al., 2021), Research and
Creative Activities Symposium 2020 (Harris et al., 2020), 2021 (Harris
et al., 2021), and 2022 (Harris et al., 2022)) with feedback collected
from participants. We have chosen the activities for this module that
proved most successful in terms of education and sparking interest.

\hypertarget{delivery}{%
\section{Delivery}\label{delivery}}

The modules are designed to be presented to participants in numeric
order as part of a workshop, skipping those marked as ``Bonus'' as
needed to fit the alloted time. Participants are encouraged to work
together to complete the modules and compare their results to one
another. While working through a module, the instructor(s) should be
available to answer questions and check in on participants' progress,
but they should leave the bulk of the work to the participants
themselves. If any bonus modules are being skipped, the instructor(s)
may wish to suggest them to participants who find they are completing
the content ahead of schedule.

All materials are designed to work on myBinder.org (Jupyter et al.,
2018), a website for hosting and interacting with Jupyter notebooks.
This is done to allow people to participate in the workshop without
needing to install any software beforehand and is treated as the default
delivery method. Participants who are experienced with Python and
already have a Jupyter environmental installed may choose instead to run
the modules locally.

\hypertarget{story}{%
\section{Story}\label{story}}

This project emerged from years of literary analysis and studying the
reproducibility of rotation curve research. This journey impressed upon
us a lack of clarity and accessibility for newcomers in the world of
rotation curves, not only in publications, but also in using software
and acquiring pre-existing data for rotation curve composition. The
problems we encountered stemmed from the resources we found being very
dense with technical language, focusing heavily on one or two components
or even parameters, or assuming the reader has a certain level of
familiarity with the subject prior to finding the resource in question.
These traits are favorable for scientific journal content, but the lack
of other types of content made it difficult to find an entry point to
the field. Our solution at the time was to dig into Noordermeer's paper
on flattened Sérsic bulges (Edo Noordermeer, 2008), a paper that took us
roughly a year to reproduce as we followed chains of references,
corresponded with authors, and tried out rotation curve construction
software in order to understand each rotation curve component. What we
hope to accomplish is to provide others with necessary vocabulary and
background knowledge before prompting them to explore this kind of
literature. Based on this experience and on feedback from our previous
workshops and presentations at research symposiums, we have developed
our own software with a focus on improving accessibility and users'
understanding of the material by being clear, concise, and easily
reproducible.

\hypertarget{acknowledgements}{%
\section{Acknowledgements}\label{acknowledgements}}

The authors would like to thank Dr.~Martin Vogelaar at Kapteyn
Astronomical Institute, Dr.~Edo Noordermeer, and Dr.~Emily E. Richards
for useful feedback on the current literature.

\hypertarget{references}{%
\section*{References}\label{references}}
\addcontentsline{toc}{section}{References}

\hypertarget{refs}{}
\begin{CSLReferences}{1}{0}
\leavevmode\hypertarget{ref-Carroll2006}{}%
Carroll, B. W., \& Ostlie, D. A. (2006). \emph{An introduction to modern
astrophysics}. Cambridge University Press.

\leavevmode\hypertarget{ref-Casertano1983}{}%
Casertano, S. (1983). {Rotation curve of the edge-on spiral galaxy NGC
5907: disc and halo masses}. \emph{Monthly Notices of the Royal
Astronomical Society}, \emph{203}(3), 735--747.
\url{https://doi.org/10.1093/mnras/203.3.735}

\leavevmode\hypertarget{ref-Epinat2008}{}%
Epinat, B., Amram, P., Marcelin, M., Balkowski, C., Daigle, O.,
Hernandez, O., Chemin, L., Carignan, C., Gach, J.-L., \& Balard, P.
(2008). {GHASP: an H\(\alpha\) kinematic survey of spiral and irregular
galaxies -- VI. New H\(\alpha\) data cubes for 108 galaxies}.
\emph{Monthly Notices of the Royal Astronomical Society}, \emph{388}(2),
500--550. \url{https://doi.org/10.1111/j.1365-2966.2008.13422.x}

\leavevmode\hypertarget{ref-Fraternali2011}{}%
Fraternali, F., Sancisi, R., \& Kamphuis, P. (2011). A tale of two
galaxies: Light and mass in NGC~891 and NGC~7814. \emph{A\&A},
\emph{531}, A64. \url{https://doi.org/10.1051/0004-6361/201116634}

\leavevmode\hypertarget{ref-Graham2001}{}%
Graham, A. W. (2001). An investigation into the prominence of spiral
galaxy bulges. \emph{The Astronomical Journal}, \emph{121}(2), 820--840.
\url{https://doi.org/10.1086/318767}

\leavevmode\hypertarget{ref-RaCAS2020}{}%
Harris, K., Bergfalk, J., \& Hatami, R. (2020). \emph{Visualizing the
evidence for dark matter}. The University of Colorado Denver.
\url{https://sites.google.com/view/racas2020/natural-physical-sciences/n15-harris-hatami-bergfalk?authuser=0}

\leavevmode\hypertarget{ref-RaCAS2021}{}%
Harris, K., Bergfalk, J., Vititoe, F., \& Hatami, R. (2021). \emph{Could
black holes explain dark matter?} The University of Colorado Denver.
\url{https://symposium.foragerone.com/2021-racas/presentations/26710}

\leavevmode\hypertarget{ref-RaCAS2022}{}%
Harris, K., Bergfalk, J., Vititoe, F., Hatami, R., \& Johnston, J.
(2022). \emph{Dark matter workshop}. The University of Colorado Denver.
\url{https://www.ucdenver.edu/sites/research-day/event-details\#ac-special-session-2-dark-matter-workshop-0}

\leavevmode\hypertarget{ref-Jimenez2003}{}%
Jimenez, R., Verde, L., \& Oh, S. P. (2003). {Dark halo properties from
rotation curves}. \emph{Monthly Notices of the Royal Astronomical
Society}, \emph{339}(1), 243--259.
\url{https://doi.org/10.1046/j.1365-8711.2003.06165.x}

\leavevmode\hypertarget{ref-Binder}{}%
Jupyter, Project, Bussonnier, Matthias, Forde, Jessica, Freeman, Jeremy,
Granger, Brian, Head, Tim, Holdgraf, Chris, Kelley, Kyle, Nalvarte,
Gladys, Osheroff, Andrew, Pacer, M., Panda, Yuvi, Perez, Fernando,
Ragan-Kelley, Benjamin, \& Willing, Carol. (2018). {B}inder 2.0 -
{R}eproducible, interactive, sharable environments for science at scale.
In Fatih Akici, David Lippa, Dillon Niederhut, \& M. Pacer (Eds.),
\emph{{P}roceedings of the 17th {P}ython in {S}cience {C}onference} (pp.
113--120). \url{https://doi.org/10.25080/Majora-4af1f417-011}

\leavevmode\hypertarget{ref-Gipsy1992}{}%
Kapteyn Astronomical Institute. (1992). \emph{GIPSY, the GRONINGEN image
processing system}. \url{https://www.astro.rug.nl/g̃ipsy/}

\leavevmode\hypertarget{ref-Karukes2015}{}%
Karukes, E. V., Salucci, P., \& Gentile, G. (2015). The dark matter
distribution in the spiral NGC 3198 out to 0.22 rvir. \emph{A\&A},
\emph{578}, A13. \url{https://doi.org/10.1051/0004-6361/201425339}

\leavevmode\hypertarget{ref-Loizides2016}{}%
Kluyver, T., Ragan-Kelley, B., Pérez, F., Granger, B., Bussonnier, M.,
Frederic, J., Kelley, K., Hamrick, J., Grout, J., Corlay, S., Ivanov,
P., Avila, D., Abdalla, S., Willing, C., \& team, J. development.
(2016). Jupyter notebooks - a publishing format for reproducible
computational workflows. In F. Loizides \& B. Scmidt (Eds.),
\emph{Positioning and power in academic publishing: Players, agents and
agendas} (pp. 87--90). IOS Press.
\url{https://eprints.soton.ac.uk/403913/}

\leavevmode\hypertarget{ref-Sparc2016}{}%
Lelli, F., McGaugh, S. S., \& Schombert, J. M. (2016). SPARC: Mass
models for 175 disk galaxies with SPITZER photometry and accurate
rotation curves. \emph{The Astronomical Journal}, \emph{152}(6), 157.
\url{https://doi.org/10.3847/0004-6256/152/6/157}

\leavevmode\hypertarget{ref-Zeropoint2015}{}%
Mamajek, E. E., Torres, G., Prsa, A., Harmanec, P., Asplund, M.,
Bennett, P. D., Capitaine, N., Christensen-Dalsgaard, J., Depagne, E.,
Folkner, W. M., Haberreiter, M., Hekker, S., Hilton, J. L., Kostov, V.,
Kurtz, D. W., Laskar, J., Mason, B. D., Milone, E. F., Montgomery, M.
M., \ldots{} Stewart, S. G. (2015). {IAU 2015 Resolution B2 on
Recommended Zero Points for the Absolute and Apparent Bolometric
Magnitude Scales}. \emph{arXiv e-Prints}, arXiv:1510.06262.
\url{http://arxiv.org/abs/1510.06262}

\leavevmode\hypertarget{ref-Skyview1998}{}%
McGlynn, T., Scollick, K., \& White, N. (1998). {SKYVIEW:The
Multi-Wavelength Sky on the Internet}. In B. J. McLean, D. A. Golombek,
J. J. E. Hayes, \& H. E. Payne (Eds.), \emph{New horizons from
multi-wavelength sky surveys} (Vol. 179, p. 465).

\leavevmode\hypertarget{ref-de_Naray2008}{}%
Naray, R. K. de, McGaugh, S. S., \& Blok, W. J. G. de. (2008). Mass
models for low surface brightness galaxies with high-resolution optical
velocity fields. \emph{The Astrophysical Journal}, \emph{676}(2),
920--943. \url{https://doi.org/10.1086/527543}

\leavevmode\hypertarget{ref-de_Naray2006}{}%
Naray, R. K. de, McGaugh, S. S., Blok, W. J. G. de, \& Bosma, A. (2006).
High-resolution optical velocity fields of 11 low surface brightness
galaxies. \emph{The Astrophysical Journal Supplement Series},
\emph{165}(2), 461--479. \url{https://doi.org/10.1086/505345}

\leavevmode\hypertarget{ref-Newville2021}{}%
Newville, M., Otten, R., Nelson, A., Ingargiola, A., Stensitzki, T.,
Allan, D., Fox, A., Carter, F., Michał, Osborn, R., Pustakhod, D.,
lneuhaus, Weigand, S., Glenn, Deil, C., Mark, Hansen, A. L. R.,
Pasquevich, G., Foks, L., \ldots{} Persaud, A. (2021).
\emph{Lmfit/lmfit-py: 1.0.3} (Version 1.0.3) {[}Computer software{]}.
Zenodo. \url{https://doi.org/10.5281/zenodo.5570790}

\leavevmode\hypertarget{ref-Noordermeer2008}{}%
Noordermeer, Edo. (2008). {The rotation curves of flattened Sérsic
bulges}. \emph{Monthly Notices of the Royal Astronomical Society},
\emph{385}(3), 1359--1364.
\url{https://doi.org/10.1111/j.1365-2966.2008.12837.x}

\leavevmode\hypertarget{ref-NoordermeerHulst2007}{}%
Noordermeer, E., \& Van Der Hulst, J. M. (2007). {The stellar mass
distribution in early-type disc galaxies: surface photometry and
bulge--disc decompositions}. \emph{Monthly Notices of the Royal
Astronomical Society}, \emph{376}(4), 1480--1512.
\url{https://doi.org/10.1111/j.1365-2966.2007.11532.x}

\leavevmode\hypertarget{ref-Richards2015}{}%
Richards, E. E., Zee, L. van, Barnes, K. L., Staudaher, S., Dale, D. A.,
Braun, T. T., Wavle, D. C., Calzetti, D., Dalcanton, J. J., Bullock, J.
S., \& Chandar, R. (2015). {Baryonic distributions in the dark matter
halo of NGC 5005}. \emph{Monthly Notices of the Royal Astronomical
Society}, \emph{449}(4), 3981--3996.
\url{https://doi.org/10.1093/mnras/stv568}

\leavevmode\hypertarget{ref-Rubin1978}{}%
Rubin, V. C., Ford, Jr., W. K., \& Thonnard, N. (1978). {Extended
rotation curves of high-luminosity spiral galaxies. IV. Systematic
dynamical properties, Sa -\textgreater{} Sc.} \emph{225}, L107--L111.
\url{https://doi.org/10.1086/182804}

\leavevmode\hypertarget{ref-Taylor1996}{}%
Taylor, J. R. (1996). \emph{An introduction to error analysis: The study
of uncertainties in physical measurements} (2 Sub). University Science
Books. ISBN:~\href{https://worldcat.org/isbn/093570275X}{093570275X}

\leavevmode\hypertarget{ref-Megaparsec}{}%
Technology, S. U. of. (n.d.). \emph{Megaparsec: cosmos}.
\url{https://astronomy.swin.edu.au/cosmos/m/megaparsec}

\leavevmode\hypertarget{ref-ScientificAmerican2008}{}%
The smallest known black hole. (2008). In \emph{Scientific American}.
Springer Nature.
\url{http://www.scientificamerican.com/gallery/the-smallest-known-black-hole}

\leavevmode\hypertarget{ref-Datathief2006}{}%
Tummers, B. (2006). \emph{Software. DataThief: Vol. NA}.
\url{https://datathief.org}

\leavevmode\hypertarget{ref-DataScienceSymposium2021}{}%
Villano, A. N., Bergfalk, J., Hatami, R., Harris, K., Vititoe, F., \&
Johnston, J. (2021). \emph{Data science symposium: The data behind dark
matter: Exploring galactic rotation}. The University of Colorado Denver.
\url{https://datascience.ucdenver.edu/events/symposium/research-session-talks}

\leavevmode\hypertarget{ref-Villano2022}{}%
Villano, A. N., Bergfalk, J., Hatami, R., Harris, K., Vititoe, Francis,
\& Johnston, J. (2022). \emph{{The Data Behind Dark Matter: Exploring
Galactic Rotation {[}Code, v1.0.1{]}}} (Version v1.0.2) {[}Computer
software{]}. The code can be found under
\url{https://github.com/villano-lab/galactic-spin-W1/}. Zenodo.
\url{https://doi.org/10.5281/zenodo.6588350}

\leavevmode\hypertarget{ref-NASA_planets2021}{}%
Williams, D. (2021, December). \emph{Planetary fact sheet}. NASA.
\url{https://nssdc.gsfc.nasa.gov/planetary/factsheet/}

\end{CSLReferences}

\end{document}